\documentclass[12pt,a4paper,reqno]{amsart}
\usepackage{graphicx} 
\usepackage{amsmath,amssymb,amsthm,array}
\usepackage{amsfonts, amsmath, amsthm, amssymb} 

\newcommand{\sech}{\mathop{\mathrm{sech}}\nolimits}

\begin{document}
\title[On perturbations retaining conservation laws]{On perturbations retaining conservation laws of difftrential equations}

\author{Alexey Samokhin}\vspace{6pt}

\address{Institute of Control Sciences of Russian Academy of Sciences
65 Profsoyuznaya street, Moscow 117997, Russia}\vspace{6pt}

\email{ samohinalexey@gmail.com}\vspace{6pt}

\begin{abstract}
The paper deals with perturbations  of the equation that have a number of conservation laws. When a small term is added to the equation its conserved quantities usually decay at individual rates, a phenomenon  known as a selective decay. These rates are described by the simple law using the conservation laws' generating functions and the added term. Yet some perturbation may retain a specific quantity(s), such as energy, momentum and other physically important characteristics of solutions. We introduce a procedure for finding such  perturbations and demonstrate it by  examples including the KdV-Burgers equation and a system from magnetodynamics. Some interesting properties of solutions of such perturbed equations are revealed and discussed.\vspace{3mm}

\noindent\textbf{Keywords:} conservation laws, perturbed equations, selective decay, traveling waves.

\noindent\textbf{MSC[2010]:} 35Q53, 35B36.
\end{abstract}

\maketitle

\section{Introduction}

Many physical systems are modeled using equations that have a significant number of conservation laws. Yet
when an additional (usually dissipative) term is added to the equation its conserved quantities decay at individual rates, which are connected to their generating functions. The famous example is the KdV equation (it has infinitely many conservation laws) and the KdV-Burgers equation (with additional, with respect to KdV, dissipative term and only one conservation law).

Namely, let ${\mathbf{E}({\mathbf u})}=0 $ be a system of
equations describing an ideal (unperturbed) media state. A scalar
$H$ depending on ${\mathbf u}$ and its derivatives is a conserved quantity  if for
 $\langle H\rangle$, the integral of $H$ over some fixed spatial domain,
$\frac{\partial \langle H\rangle}{\partial t}\bigg\vert_{\mathbf E}=0$.

For the perturbed equation the quantity $H$ is constant no more
and $\frac{\partial \langle H\rangle}{\partial t}\ne 0$
 is called the decay rate
of $H$, cf. \cite{3}.

A perturbed state usually satisfies the equation
 ${\mathbf E({\mathbf u})}+\mathfrak{L}{\mathbf F({\mathbf u})}=0$, where $\mathfrak{L}$
is a small-parameter diagonal matrix $\mathrm{diag}(\lambda_i)$; for
$\mathfrak{L}=0$ we get the ideal
state equation. The decay rate depends on the
additional term $\mathfrak{L}\, {\mathbf F(}\mathbf u)$. The connection
 between decay rate
and $\mathfrak{L}{\mathbf F({\mathbf u})}$ was called a 'balance law' in \cite{1}.

 This law expresses $\partial_t \langle H\rangle$ in terms
of scalar product of $\mathfrak{L}{\mathbf F({\mathbf u})}$ and the
generating function ${\mathbf g}$ of the conserved quantity $H$, \cite{sam1}:

\begin{equation}\label{1.1}
\frac{\partial \langle H\rangle}{\partial t}= \langle
{\mathbf g}\cdot\mathfrak{L}{\mathbf F}\rangle
\end{equation}%
\textbf{Remarks}

\begin{itemize}
  \item The right-hand side of \eqref{1.1} is not unique: e.g, one can get a different but equivalent form adding an exact form or by integrating by parts.
  \item In the case of the integrand in the right-hand side of \eqref{1.1} is null or an exact form we get the situation when the conserved quantity $\langle H\rangle$ is conserved  for the perturbed state as well.
  \item Let us restrict considerations to $\mathfrak{R}[\mathbf{u}]$, the ring of differential polynoms of $\mathbf{u}$. Then  all perturbations $\mathbf{F}$ retaining the conservation law with the generating function $\mathbf g$ must satisfy
      \[({\mathbf g}\cdot\mathfrak{L}{\mathbf F})dx_1\dots dx_n\in \mathrm{Im}(d),
      \]
where $d:\Lambda^{n-1}\rightarrow \Lambda^n$ and $\Lambda^k$ are differential $k$-forms of spatial variables.
Of course, the intersection of the principal ideal $\mathbf{g}\cdot \mathfrak{R}[\mathbf{u}]$ with $\mathrm{Im}(d)$ is huge.

\end{itemize}


 A considerable
difference in decay rates leads to a simple method, first discovered by
Taylor, \cite{5}, for finding quasi-stationary states of plasma which are of
great practical importance.

He studied the model where the decay of energy $E$ is monotonic but those of momentum $M$ and
helicity are not necessarily so. It leads to a
distinct physical phenomenon of 'self--organization' or quasi--stable states.

A simple procedure for finding solutions of such a
behavior was suggested in \cite{5}, and is known as 'Taylor trick'.
 The procedure is as follows.

Taking into consideration their comparative decay rates,
 minimize $E$ with $M$  as constrain. Put $\delta(E+\lambda
M=0)$, $M$ and presumed constant, $\lambda$  being
Lagrange multiplier.
This Euler--Lagrange equation
is not necessarily compatible with the initial equation but nevertheless it leads to  good approximations of self-organization phenomena. For instance, these approximation lead to stable numerical modelling

There is a considerable number of publication in the field, see a recent paper \cite{11} for new developments.

Another application of selective decay is given in \cite{sam2}. The problem is the behavior of the soliton  which, while moving in non-dissipative and dispersion-constant medium encounters a finite-width barrier with varying  dissipation and/or  dispersion; beyond the layer dispersion is constant (but not necessarily of the same value)  and dissipation is null. The transmitted wave either retains the form of a soliton (though of different parameters) or scatters a into a number of them. Using the relative decay of the KdV conserved quantities inside the nonhomogeneous media
 a very simple algorithm to predict the number and amplitudes of resulting solitons was obtained.

  In \cite{sam3} the selective decay approach was  applied to some well-known equations of mathematical physics (KdV and KdV-Burgers equation, BBM and its dissipative generalization, two-dimensional generalized shallow water wave equation). It have showed that the Taylor trick  extremals are associated with first-order PDEs and travelling wave solutions.

  In this paper we search, for some popular equations, their low-order
   perturbations which retain a chosen conservation law (in a sense that the perturbed equation has the same conserved quantity as initial one). Examples include KdV and its conserved energy or momentum and the Kadomtsev-Pogutse  system of equation from magnetohydrodynamics with its three known conserved quantities. Some interesting properties of solutions of such perturbed equations are revealed and discussed.

\section{KdV and KdV-Burgers}

 The  generalized KdV equation (KdV-Burgers equation) considered here is of the form
 \begin{equation}\label{01}
    u_t=2uu_x+u_{xxx}+\lambda u_{xx};
    \end{equation}
 The classical KdV equation corresponds to $\lambda=0$.

The first three conserved quantities for KdV are

\begin{equation}\label{p}
  \begin{array}{ccl}
  m = \int_{-\infty}^{+\infty}u(x,t)\,dx \mbox{               --- mass,}\\
  M = \int_{-\infty}^{+\infty}u^2(x,t)\,dx  \mbox{            --- momentum,}\\
E = \int_{-\infty}^{+\infty}\left(2u^3(x,t)-3(u_x(x,t))^2\right)\,dx  \mbox{  --- energy,}
\end{array}
\end{equation}
and there are infinite number of them.

The generating functions for the above conservation laws of the KdV are, up to multiplication constants, $1$, $u$ and $u^2+u_{xx}$ correspondingly.

As for the equation \eqref{01}, it has a form of a conservation law, $u_t=F_x$, the "mass" $\displaystyle{\int_{-\infty}^{+\infty}u\,dx}$ is a conserved quantity. For a soliton this mass is equal to $12a\gamma$.

But the impulse $\langle u^2\rangle=\displaystyle{\int_{-\infty}^{+\infty} u^2\, dx}$ declines monotonically:

\begin{equation}\label{e2}
  \begin{array}{cc}
  M_t=\frac{1}{2} \langle u^2\rangle_t=
   \langle uu_t\rangle=
   \langle u(u^2+u_{xx}+\lambda u_{x})_x\rangle 
    \left.
    =\frac{2}{3}u^3\right|_{-\infty}^{+\infty} -\left. u_x^2\right|_{-\infty}^{+\infty} -\lambda\langle u^2_{x}
\rangle  &=  \\[3mm]
  \end{array}
\end{equation}

By analogy, for the energy

\begin{equation}\label{e}
 E_t= \langle \left(2u^3(x,t)-3(u_x(x,t))^2\right)\rangle_t=6\lambda \langle u_{xx}(u^2+u_{xx})\rangle
\end{equation}

Thus the energy does not necessary declines.

\subsection{Transformations of KdV that retain momentum}

Now let us  find perturbations of the form $F(u,u_x, u_{xx})$ that retain momentum. Accordingly to the remark 2 above, the differential form $\lambda u\cdot F(u,u_x, u_{xx})dx$ must be exact. Thus

\begin{equation}\label{k1}
 u\cdot F(u,u_x, u_{xx})=D_x(A(u,u_x))
\end{equation}
for some $A(u,u_x)$. Here
\[D_x=\frac{\partial}{\partial x} +\sum_{n=0}^\infty u_{x^{n+1}}\frac{\partial}{\partial u_{x^n}}
\]
is the operator of the full differentiation with respect to $x$.

 Below we restrict the search to  polynomials  of $u$ and its derivatives. Then in \eqref{k1} the polynomial $D_x(A(u,u_x))$ is divisible by $u$, so $A(u,u_x)=u^2B(u,u_x)$.

 On the other hand
\[D_x(u^2B(u,u_x))=2uu_xB(u,u_x)+u^2(u_x\frac{\partial B}{\partial u}+u_{xx}\frac{\partial B}{\partial u_{x}}).
\]
Hence the second order retaining momentum perturbation is defined by

\[F(u,u_x, u_{xx})=2u_xB(u,u_x)+u(u_x\frac{\partial B}{\partial u}+u_{xx}\frac{\partial B}{\partial u_{x}})
\]
for an arbitrary $B$. Note that $F$ is linear in $u_{xx}$.

For instance, if $B =u_x$ the $\lambda$ transformation of the KdV equation
\begin{equation}\label{02}
    u_t=2uu_x+u_{xxx}+\lambda (2u_x^2+ uu_{xx})
    \end{equation}
retains $\langle u^2\rangle$  as its conserved quantity.

\textbf{Remark 1.} This construction can be generalized. If $g$ is the generating function for some conserved quantity $Cl$ of an one-spational equation $E$, then $F=g^{-1}D_x(g^2 \Phi)$ is the addendum to $E$ which retains  $Cl$, $\Phi$ being  a arbitrary function of $u$ and its derivatives.

\textbf{Remark 2.} The equation \eqref{02} has travelling wave solutions, in particular shock waves of the form

\begin{equation}\label{03}
 \frac{3}{2\lambda} \left(
 a\tanh\left(
 \frac{a^3\lambda^2+3a}{\lambda^2}t+ax
 \right)+\frac{1}{\lambda}
 \right).
\end{equation}

\begin{figure}[h]

\includegraphics[width=0.45\textwidth]{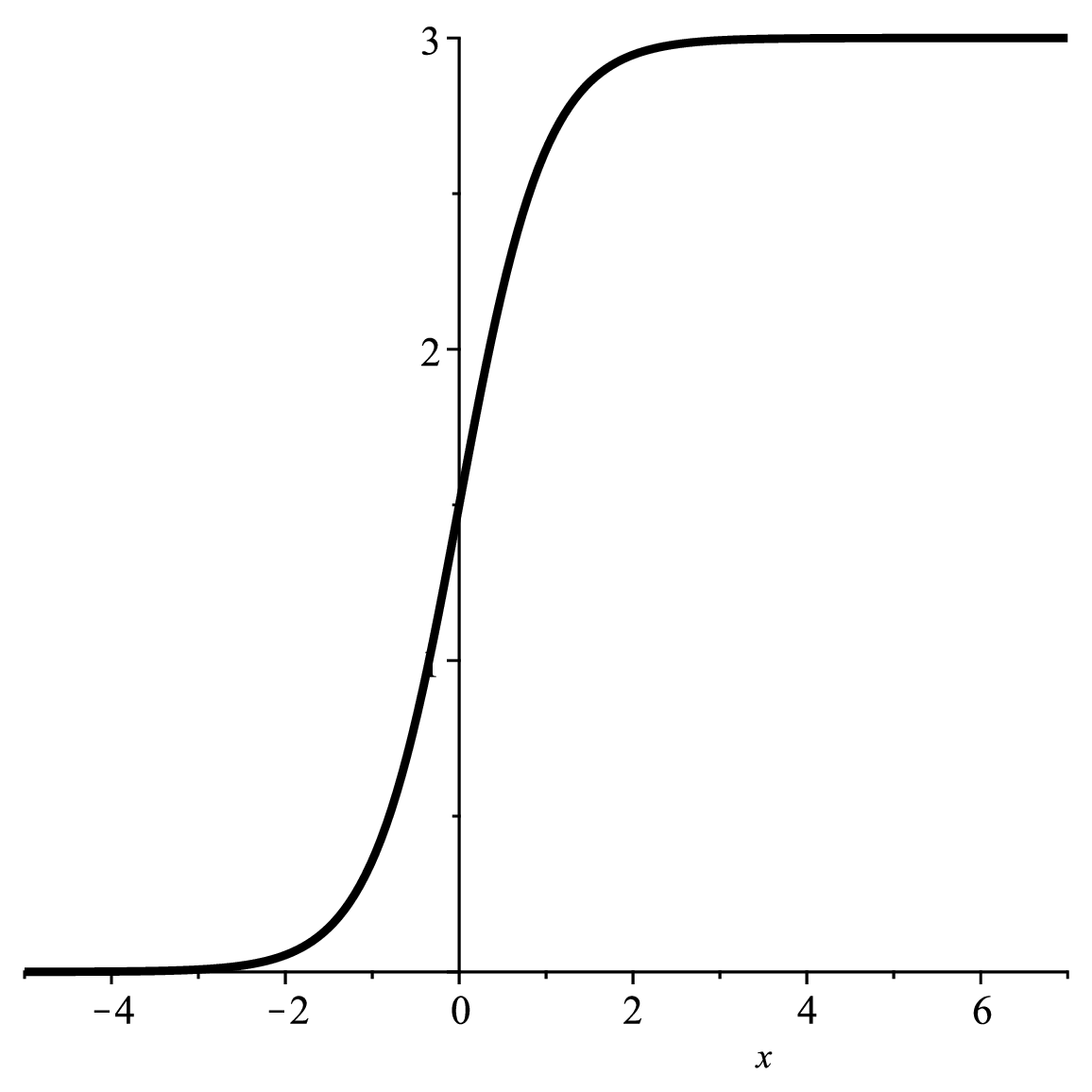}
\caption{The travelling wave solution of the equation \eqref{02}, $\lambda=1$}
\label{111}
\end{figure}

This shock  moves to the left. If require $u|_{-\infty}=0$ then \eqref{03} becomes the shock wave

\[\frac{3}{2\lambda^2} \left(1+\tanh\left(\frac{4}{\lambda^2}t+\frac{1}{\lambda}x\right)\right)
\]
with the velocity $4/\lambda$, see figure \ref{111}.

\begin{figure}[h]

\includegraphics[width=0.45\textwidth]{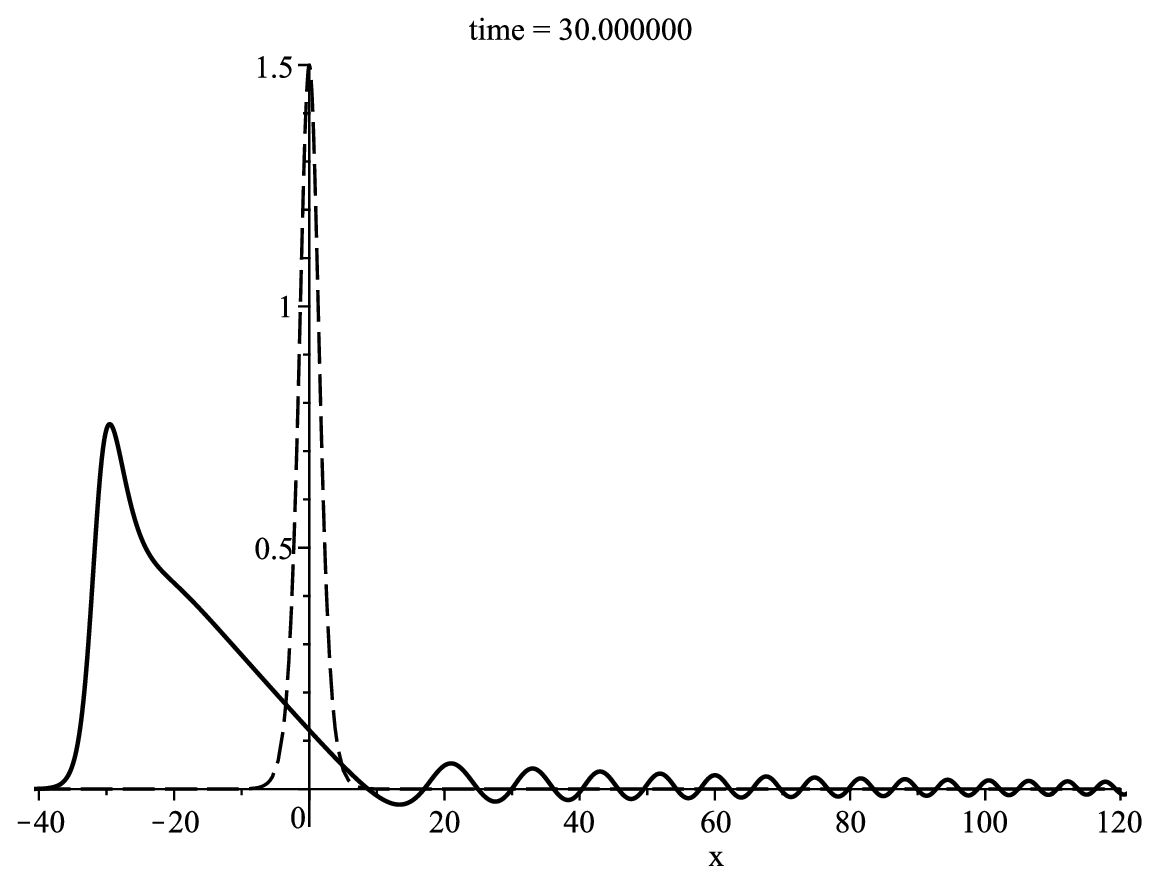}
\caption{Shock front at $t=30$ (solid line) of the decaying initial profile $ 1.5\sech^2(0.5 x)$ (dash line) for  the equation \eqref{02}, $\lambda=1$,}
\label{shock}
\end{figure}

\begin{figure}[h]

\includegraphics[width=0.45\textwidth]{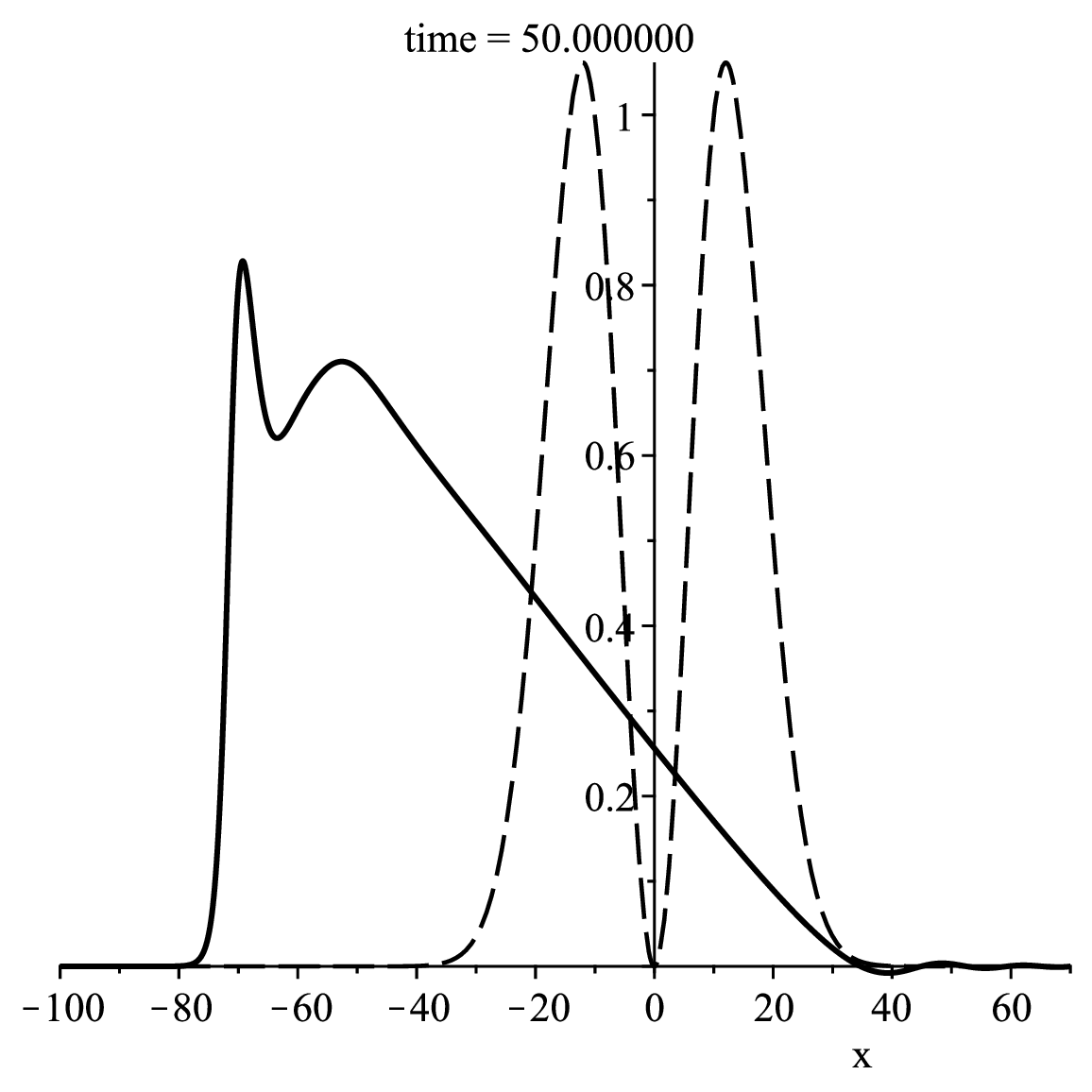}
\caption{Shock front at $t=50$ (solid line) of the decaying initial profile$(0.1x)^2 2^{1-(0.1x)^2}$ (dash line) for  the equation \eqref{02}, $\lambda=1$,}
\label{shock2}
\end{figure}

Moreover, our computer experiments show that an initial compact profile becomes a decreasing (in an amplitude and velocity)  shock wave with an oscillating tail just in the manner of the KdV-Burgers equation, see figures \ref{shock} and \ref{shock2}.
,

\textbf{Remark 4.} The perturbed equation has only translations in $x$ and $t$ as its point symmetries, but a lot of conservation laws.

\subsection{Transformations of KdV that retain energy}

Now for energy saving transformations of KdV. Since the generating function of energy is, up to a constant multiplier, $u^2+u_{xx}$, one must solve
\begin{equation}\label{k2}
 (u^2+u_{xx})\cdot F(u,u_x, u_{xx}, u_{xxx})=D_x(A(u,u_x,u_{xx}))
\end{equation}
for some $A(u,u_x,u_{xx})$, to find an low-order $F(u,u_x, u_{xx})$, the suitable transformation term. By analogy to the momentum case, the one possibility is $A=(u^2+u_{xx})^2B$

\[F(u,u_x, u_{xx})=2D_x(u^2+u_{xx})B+(u^2+u_{xx})(u_x\frac{\partial B}{\partial u}+u_{xx}\frac{\partial B}{\partial u_{x}}+u_{xxx}\frac{\partial B}{\partial u_{xx}}),
\]
for an arbitrary $B=B(u,u_x,u_{xx})$.
If $B=u$ then $F=5u^2u_x+2uu_{xxx}+u_xu_{xx}$

The corresponding transformed equation is

\begin{equation}\label{04}
    u_t=2uu_x+u_{xxx}+\lambda(5u^2u_x+2uu_{xxx}+u_xu_{xx}).
    \end{equation}
    Its point symmetries are only translations in $x$ and $t$.

\begin{figure}[h]

\includegraphics[width=0.45\textwidth]{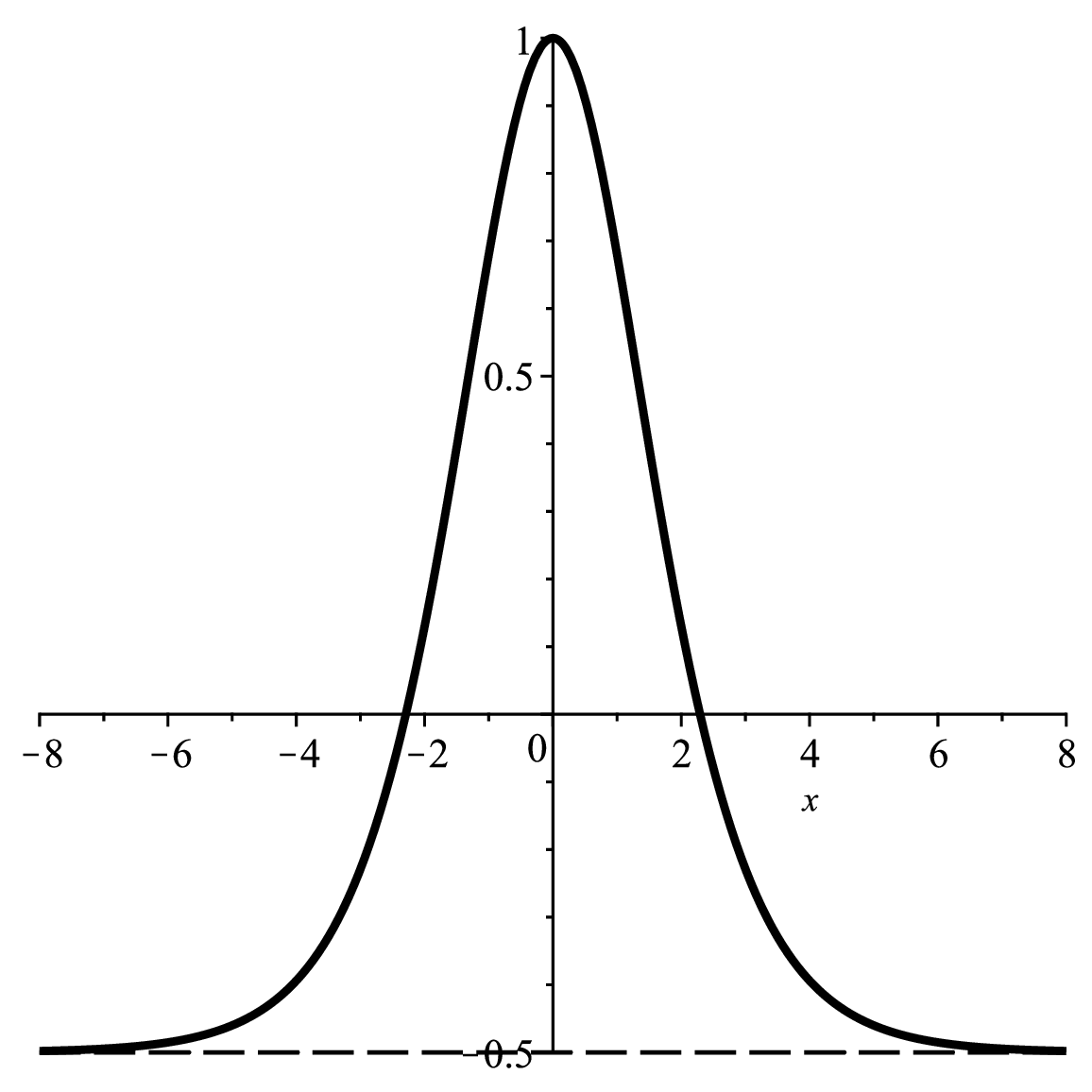}
\caption{The travelling wave soliton of the equation \eqref{04}; $\lambda=1,\; a=1/2$}
\label{222}
\end{figure}

    \textbf{Remark 5.} The equation \eqref{04} has travelling wave solutions, in particular --- solutons  of the form of a vertically shifted soliton

    \begin{equation}\label{05}
    u(x,t)=-6a^2\tanh^2(a(4a^4\cdot\lambda t+x))+4a^2= 6a^2\sech^2(a(4a^4\cdot\lambda t+x))-2a^2
\end{equation}
found by Maple, with the velocity $V=4a^4\lambda$,
see figure \ref{222}.

\begin{figure}[h]

\includegraphics[width=0.45\textwidth]{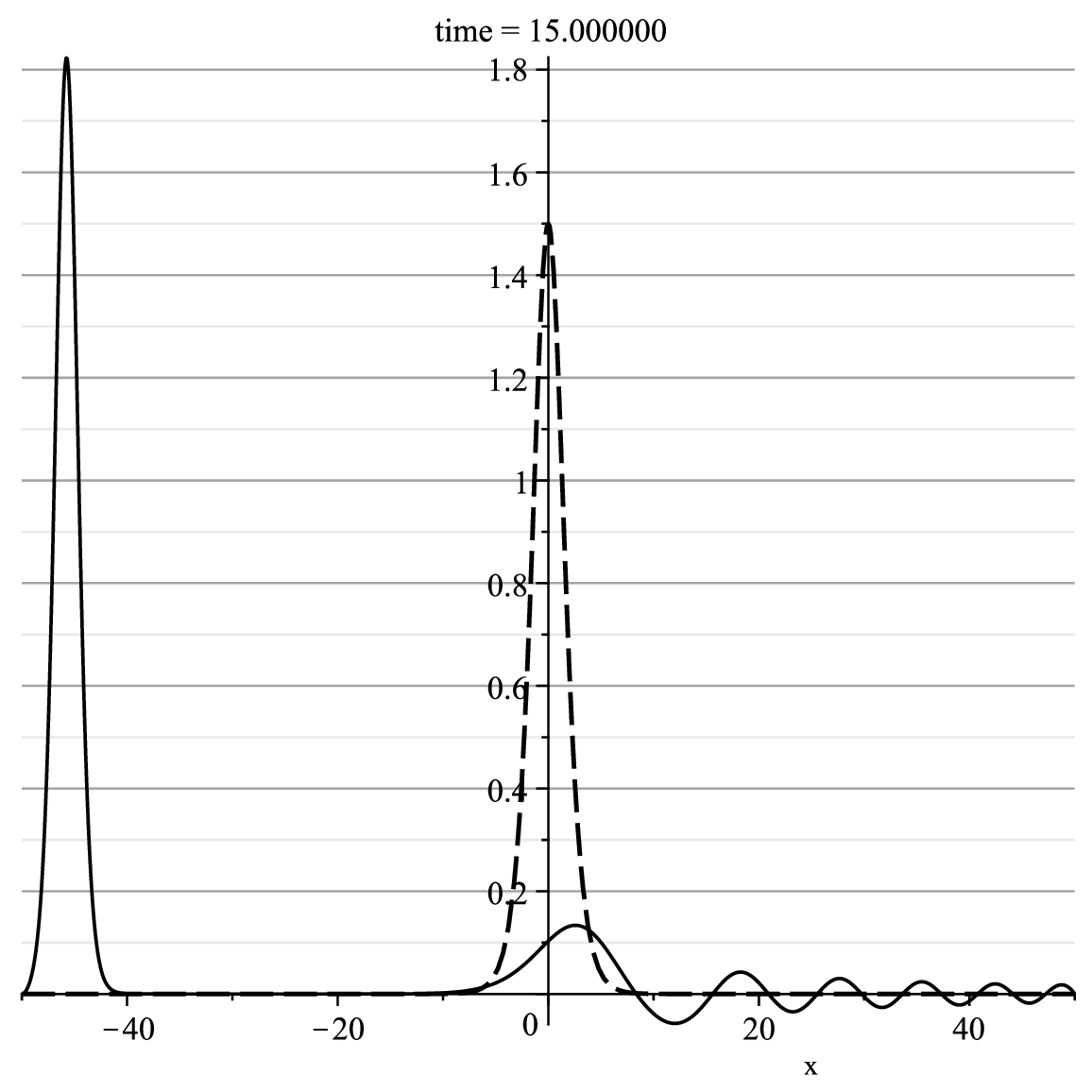}
\caption{Resulting profile at $t=15$ (solid line). corresponding to the initial one$ 1.5\sech^2(0.5 x)$ (dash line) .\\
Single soliton-like peak of a constant form and velocity and an oscillating tail moving in opposite direction}
\label{22}
\end{figure}

\begin{figure}[h]

\includegraphics[width=0.45\textwidth]{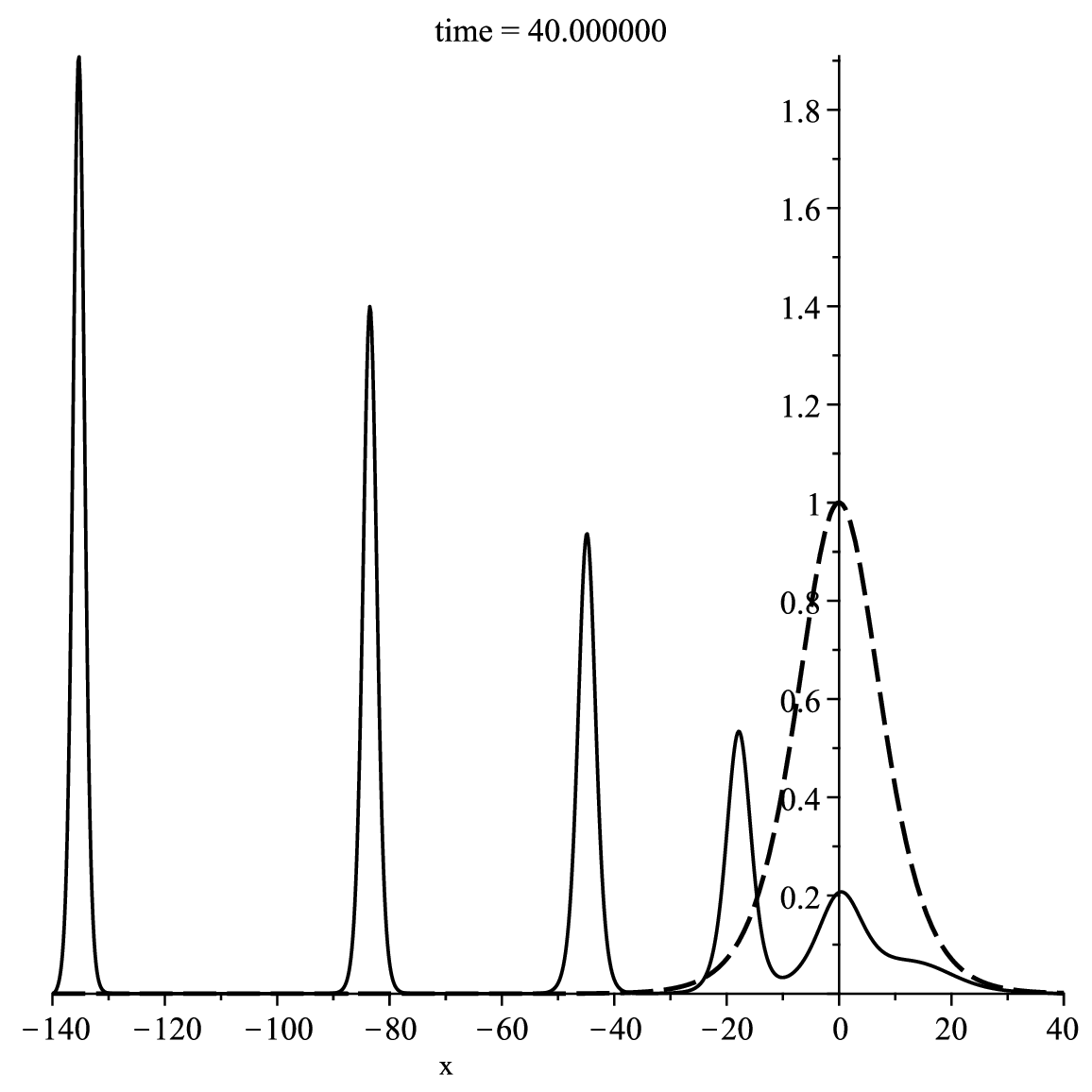}
\caption{Resulting profile at $t=40$ (solid line). corresponding to the initial one $ \sech^2(0.1 x)$ (dash line)  at $t=40$.\\
Multiple peaks (probably a series) of constant forms and velocities; seemingly, no tail.}
\label{44}
\end{figure}

Yet it is not the whole answer. Computer experiments demonstrate that an arbitrary initial datum for this equation scatters into a number of solitary peaks of different but constant height and velocity and a 'tail' (see figures \ref{22} and \ref{44}) --- in a manner of the KdV itself, cf.\cite{sam2}.

The analytical description of these peaks is so far unknown. The reason is that the equation on travelling waves, $u=u(x+Vt)$, here

\[Vu^\prime=2uu^\prime+u^{\prime\prime}+\lambda(5u^2u^\prime+2uu^{\prime\prime\prime}+u^\prime u^{\prime\prime}
\]
can be readily  integrated introducing the new dependent variable $u^\prime =p(u)$ which leads to a linear first order ordinary differential equation on $z(u)=p(u)p^\prime(u)$,
\[(2u\lambda+1)z^\prime+\lambda z=V-5\lambda u^2-2u.
\]
 But the resulting general solution looks hopelessly implicit. The likes of \eqref{05} arise in the case of a very special combination of the arbitrary constants entering this general solution, and such combinations are hard to discover.

 Recall that multi-soliton solutions for the KdV equation are not a simple sums of dingle solitons. The formolas for these multi-solitons are obtained via  the inverse scattering, not by the above straightforward integration.

\section{Two-dimensional MHD System}

Consider the Kadomtsev-Pogutse sysnem of equations
\begin{equation}\label{06}
\left\{
\begin{array}{rll}
\Delta u_t+u_x\Delta u_y-u_y\Delta u_x+v_y\Delta v_x-v_x\Delta v_y = 0 \\
v_t+u_xv_y-u_yv_x = 0
\end{array}
\right.
\end{equation}
which describes quasi-stationary states of plasma. It has three conservation laws, that is there are three non--trivial conserved densities (two of them depending on
arbitrary functions):
the total energy $E$
(magnetic plus kinetic energy), generalized 'cross helicity' $H_c$ and
mean magnetic potential $A$,

\begin{equation}\label{07}
\begin{array}{ccc}
                           E = \frac{1}{2}\langle u^2_x+u^2_y+v^2_x+v^2_y\rangle\\
                           H = \langle f^\prime(v)\cdot (u_xv_x+u_yv_y)\rangle  \\
                           A = \langle\Phi (v)\rangle
                         \end{array}
\end{equation}

Their generating functions are, respective order,
\begin{equation}\label{08}
\begin{pmatrix} u\\ \Delta v\end{pmatrix},\quad
\begin{pmatrix} f(v)\\ f^\prime(v)\Delta u\end{pmatrix},\quad
\begin{pmatrix} 0\\ \Phi^\prime (v)\end{pmatrix}
\end{equation}
where  $f$ and $\Phi$ are arbitrary functions.

Let us seek  transformations of \eqref{06} of the form

\begin{equation}\label{09}
\left\{
\begin{array}{rll}
\Delta u_t+u_x\Delta u_y-u_y\Delta u_x+v_y\Delta v_x-v_x\Delta v_y = \nu F(u,v) \\
v_t+u_xv_y-u_yv_x = \eta G(u,v)
\end{array}
\right.
\end{equation}
Here $F,\;G$ are functions of $u(x,y,t),\, v(x,y,t)$ and their derivatives.

\subsection{Energy-retaining transformations}

In this instance $\partial\langle E\rangle/\partial t=0$ implies

\begin{equation}\label{10}
\begin{array}{r}
  (-\nu u\cdot F-\eta \Delta v\cdot G)dx\wedge dy=d(A(u,v)dy-B(u,v)dx)= \\[3mm]
  (D_x A(u,v)+D_y B(u,v))dx\wedge dy.
\end{array}
\end{equation}

There are a lot of solutions to \eqref{10}. We restrict ourselves to some low-order examples.

\subsubsection{Ortogonal transformations}

One can always get zero right hand side in equation  \eqref{10}: just put $F=\eta \Delta$ and $G=-\nu u$. The vector $(F,G)$ is orthogonal to the generating function so  $\partial\langle E\rangle/\partial t=0$. It works if the number of any system of equations is greater than one.

\subsubsection{Splitted sum transformations}
Another solution may be obtained assuming

\begin{equation}\label{11}
-\nu u\cdot F(u,v)=D_x A(u,v),\quad \eta \Delta v\cdot G(u,v)=D_y B(u,v).
\end{equation}

Here again $A,B$ are functions of $u(x,y,t),\, v(x,y,t)$ and their derivatives. This equations may be solved by analogy to the KdV case.

One of numerous solutions here is $A=\nu u^n, \quad B= \eta(\Delta v)^2$, so $F=-\nu n u^{n-2}u_x,$
$G=2\eta\Delta v_y$

\subsubsection{$\nu=\eta$ case transformations}

Take $A=Gv_x,\; B=Gv_y$. Then $uF=v_xD_xG+v_yD_yG$. For instance, choose $G=u^2$; it follows that $F=2(u_xv_x+u_yv_y).$

\subsection{Mean magnetic potential retaining transformations.}

Here $\partial\langle A\rangle/\partial t=0$ implies

\begin{equation}\label{12}
\begin{array}{r}
  (-\nu 0\cdot F-\eta \Phi^\prime(v)\cdot G)dx\wedge dy=d(A(u,v)dy-B(u,v)dx)= \\[3mm]
  (D_x A(u,v)+D_y B(u,v))dx\wedge dy.
\end{array}
\end{equation}

Thus $F$ is an arbitrary function.
Then one possible solution is
\[-\eta \Phi^\prime(v)\cdot \Phi(v)(\alpha v_x+\beta v_y)=D_x\alpha \Phi^2+D_y\beta\Phi^2, \alpha,\beta \in\mathbb{R}.
\]
That is, to retain the mean magnetic potential of \eqref{06}, its first equation  may be transformed in arbitrary way and the second one by $\eta G=-\eta  \Phi(v)(\alpha v_x+\beta v_y)$ for all
$\alpha,\beta \in\mathbb{R}$.

\subsection{Cross helicity retaining transformations.}

Here $\partial\langle H_c\rangle/\partial t=0$ implies

\begin{equation}\label{13}
-\nu f(v)\cdot F(u,v)-\eta f^\prime (v)\Delta(u)\cdot G(u,v)=D_x A(u,v)+D_y B(u,v).
\end{equation}

In the case $\eta=\nu$ it is not hard to find some suitable transformations $(F,G)$. Namely, take
\[
A=-\eta{f^2(v) f^\prime}(v)u_x,\;B=-\eta f^2(v) f^\prime(v) u_y;
\]
It follows
\[F=[2 {f^\prime}^2(v)+f(v) f^{\prime\prime }(v)](v_xu_x+u_yv_y),\;
G=f^\prime(v) f(v)\Delta u.
\]

For $f(v)=v$ it comes to
\[F=-2\eta(v_xu_x+u_yv_y);\quad G=-\eta v\Delta u.
\]

\section*{Conclusion}
The paper deals with perturbations  of the equation that have a number of conservation laws. When a small term is added to the equation its conserved quantities usually decay at individual rates, a phenomenon  known as a selective decay. These rates are described by the simple law using the conservation laws' generating functions and the added term. Yet some perturbation may retain a specific quantity(s), such as energy, momentum and other physically important characteristics of solutions. We introduced a procedure for finding such  perturbations and demonstrated it by  examples including the KdV-Burgers equation and a system from magnetodynamics.

Our worked out examples show that the perturbed equations retaining a specific conservation law frequently also retain additional algebraic properties such as travelling wave solutions or a presence of other conservation laws.

Thus the present paper as well as our previous research of the KdV solitons in nonhomogeneous media, \cite{sam2}, persuades that the selective decay approach is a valid and effective instrument to obtain qualitative approximations and estimates for behavior of solutions.

The figures in this paper were generated numerically using Maple PDETools package. The  mode of operation uses the default Euler method, which is a centered implicit scheme, and  can be used to find solutions to PDEs that are first order in time, and arbitrary order in space, with no mixed partial derivatives.

\section*{Acknowledgment}
This work was partially supported by the Russian Science Foundation grant 23-21-00390.

\end{document}